\documentclass{ws-procs9x6}
  
\def\citebk#1{[\hspace{0.9mm}\raisebox{-1.85mm}[0mm][0mm]
  {\Large\cite{#1}}\hspace{-0.1mm}]}

\newcommand{\eqb}{\begin{equation}}
\newcommand{\eqe}{\end{equation}}
\newcommand{\dmb}{\begin{displaymath}}
\newcommand{\dme}{\end{displaymath}}
\newcommand{\pd}{\partial}

\newcommand{\eab}{\begin{eqnarray}}
\newcommand{\eae}{\end{eqnarray}}
\newcommand{\ra}{\right\rangle}
\newcommand{\la}{\left\langle}
\newcommand{\e}{\mbox{e}}
\newcommand{\be}{\begin{equation}}
\newcommand{\ee}{\end{equation}}

\newcommand{\La}{\Lambda}\def\lsim{\mathrel{\raise.3ex\hbox{$<$\kern-.75em\lower1ex\hbox{$\sim$}}}}
\def\gsim{\mathrel{\raise.3ex\hbox{$>$\kern-.75em\lower1ex\hbox{$\sim$}}}}
\def\Li2{{\rm Li}_2}

\newcommand{\bmE}{\mathbf E}

\newcommand{\bmx}{\mathbf x}
\newcommand{\bmy}{\mathbf y}

\begin{document}

\title{Reshuffling the OPE: \\Delocalized Operator Expansion}

\author{R. HOFMANN }

\address{Max-Planck-Institut f\"ur Physik, \\
Werner-Heisenberg-Institut\\
F\"ohringer Ring 6\\ 
80805 M\"unchen,\\ 
Germany\\ 
E-mail: ralfh@mppmu.mpg.de}


\maketitle

\abstracts{A generalization of the operator 
product expansion for Euclidean correlators of 
gauge invariant QCD currents is presented. 
Each contribution to the 
modified expansion, which is based on a delocalized multipole expansion 
of a perturbatively determined coefficient function, 
sums up an infinite series of local operators. 
On a more formal level the delocalized operator expansion 
corresponds to an {\sl optimal} choice of 
basis sets in the dual spaces which are associated with 
the interplay of perturbative and 
nonperturbative $N$-point correlations 
in a distorted vacuum. A consequence of the delocalized 
expansion is the running of condensates 
with the external momentum. Phenomenological evidence is gathered
that the gluon condensate, often 
being the leading nonperturbative
parameter in the OPE, 
is indeed a function of
resolution. Within a model 
calculation of the nonperturbative corrections 
to the ground state energy of a heavy quarkonium system it 
is shown exemplarily that the 
convergence properties are better 
than those of the OPE. Potential applications of the 
delocalized operator expansion in view of estimates of 
the violation of local quark-hadron duality are discussed.  
}

\section{Introduction}

\subsection{Some historical remarks}

Once again: Happy Birthday, Arkady! This is a very 
lively conference that I am happy to attend.

In this talk I would like to report about work done jointly 
with Andr\'e Hoang. The subject is to formulate a framework for 
a generalization of Wilson's operator product expansion (OPE), to confront 
this framework with experiment, and to demonstrate its usefulness 
in some model calculations.    

As a warm-up I remind you of some of the 
OPE applications. The OPE is a powerful tool for the treatment of 
conformally invariant field theories where its 
convergence is explicitly shown. Unfortunately, 
this feature does not survive in realistic, four-dimensional 
field theories. Here the expansion is believed to 
be asymptotic at best.       

The construction 
of effective theories like HQET or 
of the effective Hamiltonians of electroweak theory are 
crucially relying on the OPE 
which determines effective vertices by a perturbative matching 
of amplitudes with those of the fundamental theory. 
Heavy fields are effectively 
integrated out in this procedure 
(see \citebk{Buras} and references therein). 

The OPE turned out to be important for a 
semi-perturbative derivation of the
exact $\beta$ function in $N=1$ 
super Yang-Mills theory \citebk{SV86}.    

And, finally, there is a major application in QCD sum rules 
\citebk{SVZ} where the ``theoretical side'' of a dispersion relation 
for the correlator of gauge invariant currents 
is represented in terms of an OPE. The occurence of power corrections 
to the low-order in $\alpha_s$ perturbative result 
can be argued from the summation of 
bubble graphs inducing so-called infrared 
renormalons, and formal contact to the OPE can be made \citebk{renormalon}. 
A practical version of the OPE was proposed by Shifman, Vainshtein, 
and Zakharov in \citebk{SVZ} where 
nonperturbative factors in the power 
corrections, the QCD condensates, are viewed as universal 
parameters of the QCD vacuum to be 
determined by experiment. Due to 
the universality of the condensates 
this method is rather predictive in practice.

\subsection{Problems of the OPE in QCD}

Despite its many successes in the 
framework of the QCD sum 
rule method there are some unanswered 
questions concerning the 
convergence properties of the OPE. 
Answering these questions very likely resolves 
the problems connected with local quark-hadron 
duality (see \citebk{BigiUraltsev} in references therein) 
which can be formulated as follows. 

Given analyticity of the correlator away from the 
positive real axis and invoking the optical theorem 
the discontinuity of the correlator at time-like 
momenta should yield 
the spectral function of the hadronic 
channel which corresponds to the 
QCD current in the correlator. This procedure, however, fails 
badly if a ``practical'' OPE with a 
small number of power corrections is used. 
Whereas the experimentally measured 
spectral function shows resonance wiggles at 
moderate time-like momenta a continuation of 
the OPE generates a smooth behavior 
down to small momenta. This may not harm the sum rule 
method since the relevant quantity there is a 
spectral {\sl integral}, that is, an average 
over resonances which agrees 
quite well with the average over the 
associated discontinuity of the OPE. However, 
in applications, where the local 
properties of OPE discontinuities 
are needed \citebk{Nierste}, the violation of 
local quark-hadron duality 
can induce large errors.  

Another concern, which is connected with the violation of 
local quark-hadron duality, is 
the asymptotic nature of the expansion. Asymptotic behavior can 
be argued from estimates of operator averages using the instanton gas 
approximation \citebk{SVZ} or by appealing to 
the renormalon idea. From the OPE itself, where naively an expansion in powers 
of $\La_{QCD}/Q$ is assumed, there is no way of estimating 
the critical dimension at which the expansion ceases 
to approximate. In the framework of a delocalized expansion 
it was argued in \citebk{Hofmann} that the knowledge about 
nonperturbative, gauge invariant 
correlation functions would allow for 
such an estimate if the decay of effective 
correlation length with mass dimension is sufficiently fast. 

The central theme of this talk is a nonlocal 
generalization of the OPE which has the potential to cure 
the above short-comings.
        
\section{Delocalized operator expansion}

The OPE of the correlator of a current $j$ reads
\eab
\label{corr}
&&i\int d^4x\,\mbox{e}^{iqx}\la T j(x)j(0)\ra=Q^2\sum_{d=0}\sum_{l=1}^{l_d}c_{dl}(Q^2) 
\la O_{dl}(0)\ra\,,\ \ (Q^2=-q^2>0)\,,\nonumber\\ 
&&
\eae
where $d,l$ run over the dimension and 
the field content of the local operator 
$O_{dl}$, respectively. The Wilson coefficient $c_{dl}$ is 
perturbatively calculable. The expansion (\ref{corr}) 
contains reducible chains of operators, such as
\eab
\label{redc}
&&G^2, GD^2G, GD^4G, GD^6G ...\,,\nonumber\\ 
&&\bar{q}q, \bar{q}D_\mu \gamma_\mu q, \bar{q}D^2q, \bar{q}D^2D_\mu \gamma_\mu q\,,
\eae
$D_\mu$ denotes the gauge covariant derivative. 
Here ``reducible'' refers to the fact 
that each chain can be obtained from a Taylor 
expansion of an associated gauge 
invariant 2-point correlator which is not 
decomposible into gauge invariant factors. 
Examples, which correspond to the chains 
in (\ref{redc}), are the 
gluonic field strength correlator \citebk{Dosch}
\eqb
\label{fsc}
g_{\mu\nu\kappa\lambda}(x)\equiv tr\la g^2 G_{\mu\nu}(x)\,P\e^{ig\int_0^x dz_\mu A_\mu(z)}\,
G_{\kappa\lambda}(0)P\e^{ig\int_x^0 dz_\mu A_\mu(z)}\ra
\eqe
and the bilocal quark ``condensate''
\eqb
\label{bqc}
q(x)\equiv \la \bar{q}(x)\,P\e^{ig\int_0^x dz_\mu A_\mu(z)}\,q(0)\ra\,,
\eqe
where $P$ denotes path ordering. For lattice measurements the path connecting 
the points $0$ and $x$ is taken to be a straight line. This should 
capture the largest scale $\Lambda$ in the nonperturbative decay of the correlator. The general case 
of $N$-point correlations 
will be addressed below. The effect of a delocalized 
expansion is a partial summation of such a chain in each order of the momentum 
expansion of the perturbative coefficient. This leads to an improvement of the convergence 
properties. 

Let me make this explicit for the case of 
2-point correlations in a 1-dimensional 
Euclidean world \citebk{HoHo}. The contribution to the OPE, which arises from this,  
can formally be written as 
\begin{eqnarray}
\label{mult}
\int_{-\infty}^\infty dx\,f(x)\,g(x) 
& = & 
\int dx\,dy\,f(y)\,\delta(x-y)\,g(x)
\nonumber \\[2mm] 
& = &
\int dx\,dy\,f(y)\,
\bigg[\,\sum_{n=0}^{\infty}\frac{(-1)^n}{n!}\,y^n
  \delta^{(n)}(x)\,\bigg]\,
g(x)
\nonumber\\[2mm]  & = &
\sum_{n=0}^{\infty}\bigg[\,\int dy\,f(y)\,y^n\,\bigg]\,
\left[\int dx\,\frac{(-1)^n}{n!}\delta^{(n)}(x)\,g(x)\right]\nonumber\\ 
&\equiv&\sum_{n=0}^{\infty} f_n(\infty) \, g_n(\infty)\,. 
\end{eqnarray}
Here $f(x)$ is associated with perturbative correlations whereas 
$g(x)$ describes nonlocal, nonperturbative effects. 
The integral expressions in the first and second square bracket correspond to local 
Wilson coefficients and operators of the like of (\ref{redc}), respectively. 
From (\ref{mult}) it is seen that the part of the OPE being generated by 
2-point correlations is induced by a bilinear 
form $(\cdot\,,\cdot)\equiv\int dx \cdot\times\,\cdot$. The expansion (\ref{mult}) corresponds to a specific choice of 
basis $\{\tilde{e}_n(\infty),e_n(\infty)\}$ in the associated dual space of 
square integrable functions where
\eqb
\label{basis}
\tilde{e}_n(\infty)=y^n \ \ \ \ \ \ \mbox{and} \ \ \ \ \ e_n(\infty)=\frac{(-1)^n}{n!}\,\delta^{(n)}(x)\,.
\eqe
and we have orthonormality
\eqb
\label{ortho1}
\left(\,e_n\,,\,\tilde e_m\,\right)
\, = \, 
\delta_{nm}\,.
\eqe 
From Eq.\,(\ref{mult}) it follows that 
\begin{eqnarray}
\label{multexp}
f(x) & = &
\sum_{n=0}^{\infty}
\,\bigg[\, \int dy\,f(y)\,y^n  \,\bigg]\,
\frac{(-1)^n}{n!}\,\delta^{(n)}(x)
\,.
\end{eqnarray}
This corresponds to an expansion of the perturbative factor $f(x)$ into degenerate, 
that is, zero-width multipoles.
\begin{figure}
\begin{center}
\leavevmode
\epsfxsize=9.cm
\leavevmode
\epsffile[80 25 534 344]{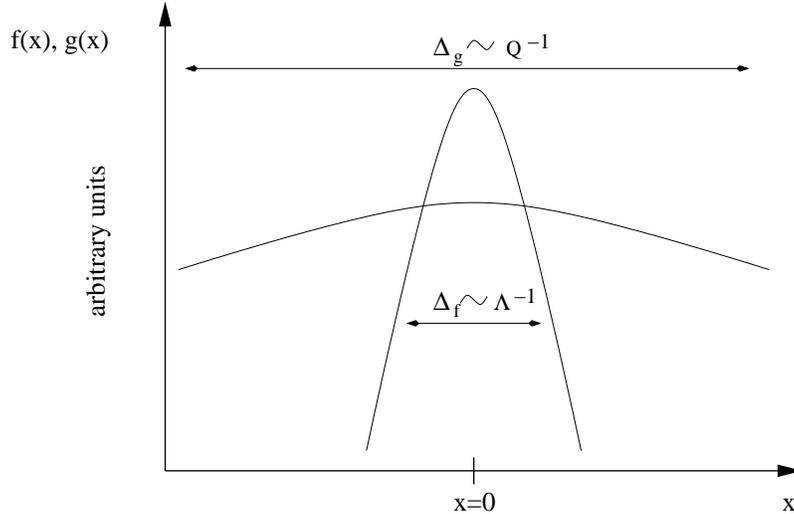}
%
\end{center}
\caption{Schematic drawing of the short-distance function $f(x)$ and
the nonperturbative 2-point-correlator $g(x)$ illustrating the scale hierarchy
$\Delta_f/\Delta_g\sim\Lambda/Q\ll 1$.
\label{figfandg}
}   
\end{figure}
It is suggestive that an expansion into multipoles of width $\sim\Omega^{-1}$, which is 
comparable with $\Delta_f\sim Q^{-1}$, converges better (see Fig.\,1). Like the OPE 
this expansion is controlled by powers of $\frac{\Delta_f}{\Delta_g}\sim \frac{\Lambda}{Q}$, and it 
breaks down if $\Delta_f\sim Delta_g$. The expansion into finite-width multipoles 
corresponds to a change of basis in dual space. Rather than expanding into the basis 
$\{\tilde{e}_n(\infty),e_n(\infty)\}$ one may expand into 
$\{\tilde{e}_n(\Omega),e_n(\Omega)\}$ where
\eqb
\label{softb}
\tilde e_n^\Omega(x)=\frac{H_n(\Omega x)}{(2 \, \Omega)^n}\,,\ \ \ 
e_n^\Omega(x)\equiv \frac{\Omega^{n+1}}{\sqrt{\pi}\,n!}\,H_n(\Omega x)\,\e^{-\Omega^2 x^2}
\eqe
We then have
\eqb
\label{profsoft}
\int dx\,f(x)\,g(x)=\sum_{n=0}^{\infty} f_n(\Omega) \, g_n(\Omega)\,.
\eqe
In (\ref{softb}) $H_n$ denotes the $n$th Hermite polynomial. An expansion into any other 
basis that allows for a sensible 
multipole expansion of $f(x)$ is possible, and the 
restriction to a Gaussian based expansion is just for definiteness. 
Note that (\ref{softb}) reduces to 
(\ref{basis}) in the limit $\Omega\to\infty$. According to (\ref{softb}) the transformations, 
which link a basis with resolution parameter $\Omega$ to a basis with $\Omega^\prime$, 
form a 1-parameter group. The moments 
$f_n(\Omega)$ in (\ref{profsoft}) have {\sl finite} expansions into the Wilson coefficients $f_n(\infty)$
\eqb
\label{fnexp}
 f_0(\Omega)=f_0(\infty)\,,\ f_2(\Omega)=f_2(\infty)-\frac{1}{2\Omega^2}f_0(\infty)\,,\cdots\,.
\eqe
Chosing a particular basis and a value for $\Omega$ is 
analogous to a choice of regularization and normalization point in pure 
perturbation theory. In practice (\ref{fnexp}) enables to consider perturbative 
renormalization and delocalized operator expansion (DOE) 
separately. The expansion 
of $g(\Omega)$ into local condensates is {\sl infinite}, for example
\eqb
\label{gnexp}
g_0(\Omega)=g_0(\infty)+\frac{1}{2\Omega^2}g_2(\infty)+\cdots\,.
\eqe
To evaluate sums like in (\ref{gnexp}) one would have to know infinitely many 
anomalous dimensions and condensates. On the other hand, one may be more pragmatic 
and start with a (lattice inspired) model for $g(x)$ which is believed to contain the information 
about all $g_n(\infty)$. To {\sl each} order $\frac{\La}{\Omega}$ one may then 
derive evolution equations for $f_n(\Omega)$ and $g_n(\Omega)$ which sum 
these contributions to {\sl all} orders \citebk{HoHo}.  

Setting $\Omega=Q$, the parametric 
organization of the expansion in $\frac{\La}{Q}$ up to an overall factor is
\eab
\label{par}
f_0(Q)\times g_0(Q)&\sim&\left(\frac{\Lambda}{Q}\right)^0\times\{\left(\frac{\Lambda}{Q}\right)^0+
\left(\frac{\Lambda}{Q}\right)^2+\cdots\}\nonumber\\ 
f_2(Q)\times g_2(Q)&\sim&\left(\frac{\Lambda}{Q}\right)^2\times\{\left(\frac{\Lambda}{Q}\right)^0+
\left(\frac{\Lambda}{Q}\right)^2+\cdots\}\nonumber\\ 
&\vdots&\ \ \ \ \ \ \ \ \ \ \ \ \ \ \ \ \ .
\eae
The generalization to $K$-dimensional Euclidean space is 
straightforward \citebk{HoHo}. It also allows for a treatment 
of $N$-point correlations in 4-dimensional space. In particular, the $n=0$ term of the 
contribution to the DOE from 2-gluon correlations yields the local Wilson coefficient times a 
``runnning gluon condensate'' ($Q=\Omega$). 

\section{Applications}

\subsection{Gluonic field strength correlator}

To have a well-motivated model for $g(x)$ for the case of 2-gluon correlations 
we appeal to a parametrization of the gluonic field strength correlator (\ref{fsc}) 
as it was proposed in \citebk{Dosch}
\eab
\label{parfsc}
g_{\mu\nu\kappa\lambda}(x)&\equiv&(\delta_{\mu\kappa}\delta_{\nu\lambda}-\delta_{\mu\lambda}\delta_{\nu\kappa})
[D(x^2)+D_1(x^2)]+\nonumber\\ 
&& (x_\mu x_\kappa\delta_{\nu\lambda}-x_\mu x_\lambda\delta_{\nu\kappa}+
x_\nu x_\lambda\delta_{\mu\kappa}-
x_\nu x_\kappa\delta_{\mu\lambda})\frac{\pd D_1(x^2)}{\pd x^2}\,.
\eae
In a lattice measurement \citebk{Delia} 
the scalar functions $D$ and $D_1$ were parametrized as
\eab
\label{latD}
D(x^2)&=&A_0\exp[-|x|/\lambda_A]+\frac{a_0}{|x|^4}\exp[-|x|/\lambda_a]\,,\nonumber\\ 
D_1(x^2)&=&A_1\exp[-|x|/\lambda_A]+\frac{a_1}{|x|^4}\exp[-|x|/\lambda_a]\,, 
\eae
where the purely exponentially decaying terms are attributed 
to purely nonperturbative dynamics. The measurements at various 
quark masses for dynamical light quarks 
indicate that for $|x|\le\lambda_A$ the nonperturbative 
contribution to $D$ dominates the nonperturbative 
contribution to $D_1$ implying that the tensor 
structure of $g_{\mu\nu\kappa\lambda}$ is identical to 
that of the corresponding local operator obtained by letting $x\to 0$. 
This feature leads to great simplifications.

\subsection{Extraction of the running gluon condensate}

Using the DOE instead of the OPE, 
the running gluon condensate can be extracted from experiment in channels 
where 2-gluon correlations are dominant. We have 
done this for the $V+A$ correlator and the charmonium system. For a more detailed 
presentation see \citebk{HoHo}. 

\subsubsection{Sum rules for the V+A correlator}

The spectral function for light quark production in the
$V+A$ channel has been remeasured recently from hadronic
$\tau$ decays for $q^2\le m_\tau^2$ by Aleph~\citebk{Aleph} 
and Opal~\citebk{Opal}. In the OPE the associated  
current correlator is dominated by the gluon condensate, and the 
dimension $6$ power corrections that are not due to a double
covariant derivative in the local gluon condensate are suppressed.  
The corresponding currents are
$j_\mu^{L/R}=\bar{u}\gamma_\mu(1\pm\gamma_5)d$ and the relevant
correlator in the chiral limit reads
\eab
i\int d^4\!x\, \e^{iqx}\la T j_\mu^{L}(x)j_\nu^{R}(0)\ra
\, &=& \,
(q_\mu q_\nu-q^2 g_{\mu\nu})\,\Pi^{\rm V+A}(Q^2)
\,,
\qquad
Q^2=-q^2
\,.\nonumber\\ 
&& 
\label{VpAdef}
\eae
Since the correlator $\Pi(Q^2)^{\rm V+A}$ is cutoff-dependent itself,
we investigate the Adler function
\begin{equation}
\label{Adler}
D(Q^2) 
\, \equiv \, 
-Q^2\,\frac{\partial\,\Pi^{\rm V+A}(Q^2)}{\partial\, Q^2}
\, = \,
\frac{Q^2}{\pi}\,\int_0^\infty ds\, 
 \frac{\mbox{Im}\,\Pi^{\rm V+A}(s)}{(s+Q^2)^2}
\,.
\label{Adlerdef}
\end{equation}
For the corresponding experimental V+A spectral function we have used
the Aleph measurement in the resonance region up to 
$2.2$~GeV$^2$. For the continuum region above $2.2$~GeV$^2$ 
we used 3-loop perturbation theory for $\alpha_s(M_Z)=0.118$, and we
have set the renormalization scale $\mu$ to $Q$.
We remark that the pion pole of the axial vector contribution has to be
taken into account in order to yield a consistent description in terms
of the OPE for asymptotically large $Q$\,\citebk{Braaten1}. 
We have checked that the known perturbative contributions to the Adler  
function show good convergence properties. For $Q$ between $1.0$ and
$2.0$~GeV and setting $\mu=Q$, the 3-loop (two-loop) corrections
amount to $5\%$ ($7\%$) and $0.5\%$ ($1.4\%$), respectively. 
On the other hand, the O$(\alpha_s)$ correction to the
Wilson coefficient of the gluon condensate are between $-16\%$ and
$-6\%$ for $\mu$ between $0.8$ and $2.0$~GeV.
We have compared the local dimension $6$ contributions contained in
the running gluon condensate with the  
sum of all dimension $6$ terms in the OPE as they were determined in 
\citebk{Braaten1}. We again found that the corresponding 
dimension $6$ contributions have equal sign and roughly the same size. 

\begin{figure}[t!] 
\begin{center}
\leavevmode
\epsfxsize=3.5cm
\epsffile[260 430 420 720]{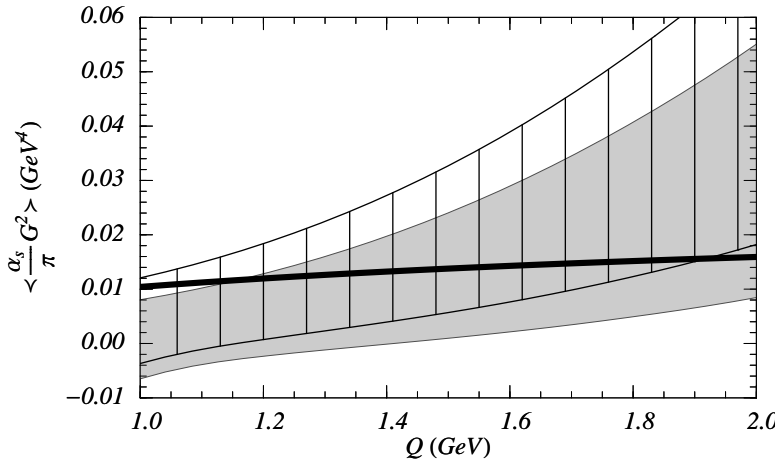}
%
%
\vspace{0cm}
 \caption{\label{figAdler} 
The running gluon condensate as a function of $Q$ when extracted from
the Adler function.
The grey area represents the allowed region using perturbation
theory at O$(\alpha_s^3)$ and the striped region using
perturbation theory at O$(\alpha_s^2)$.
The thick solid line denotes the running gluon condensate ($\Omega=Q$) from the
lattice-inspired ansatz.
}
 \end{center}
\end{figure}

Our result for the running gluon condensate as a function of $Q$ is
shown in Fig.\,\ref{figAdler}. 
for $1$\,GeV$\le Q \le 2$~GeV. The uncertainties are due to 
the experimental errors in the spectral function and a variation of the
renormalization scale $\mu$ in the range $Q\pm 0.25$~GeV.
The strong coupling has been fixed at $\alpha_s(M_Z)=0.118$. 
We have restricted our analysis to the range 
$1\,\,\mbox{GeV}<Q<2\,\,\mbox{GeV}$ because for $Q\lsim
1$~GeV perturbation theory becomes unreliable and for $Q\gsim 2$~GeV the 
experimentally unknown part of the spectral 
function at $s\ge 2.2$\,GeV$^2$ is being probed.
The thick black line in Fig.\,\ref{figAdler} shows the 
running gluon condensate obtained from the lattice-inspired model 
of the last section  for $\Omega=Q$. 
We find that our result for the running gluon condensate as a function
of $Q$ is consistent with a function that increases with $Q$.

\subsubsection{Charmonium sum rules}

The determination of the local gluon condensate 
$\Big\langle\frac{\alpha_s}{\pi}\,G^2\Big\rangle(\infty)$ from
charmonium sum rules was pioneered in Refs.\,\citebk{Novikov1}.
By now there is a vast literature on computations of Wilson
coefficients to various loop-orders and various dimensions of the 
power corrections in the
correlator of two heavy quark currents including updated
analyses of the corresponding sum rules (see
e.g. Refs.\,\citebk{Nikolaev1,Nikolaev2}).

The relevant correlator is
\begin{equation}
\label{barccc}
(q_\mu q_\nu-q^2\,g_{\mu\nu}) \,\Pi^c(Q^2) 
\, = \, 
i\int d^4x\, \e^{iqx}\la T j^c_\mu(x)j^c_\nu(0)\ra
\,,\qquad
Q^2=-q^2
\,,
\end{equation}
where $j_\mu\equiv\bar{c}\gamma_\mu c$, and the $n$-th moment is
defined as 
\begin{equation}
M_n
\, = \,
\frac{1}{n!}\,\left.\Big(\!-\frac{d}{dQ^2}\Big)^n
\,\Pi^c(Q^2)\,\right|_{Q^2=0}
\,.
\label{momdef1}
\end{equation}
Assuming analyticity of $\Pi^c$ in $Q^2$ 
away from the negative, real axis and employing 
the optical theorem, the $n$th moment can be expressed as a
dispersion integral over the charm pair cross section in $e^+e^-$
annihilation,
\begin{equation}
M_n 
\, = \,
\frac{1}{12\pi^2\,Q_c^2}\,\int \frac{ds}{s^{n+1}}\,
\frac{\sigma_{e^+e^-\to c\bar
    c+X}(s)}{\sigma_{e^+e^-\to\mu^+\mu^-}(s)}
\,,
\label{momdef2}
\end{equation}
where $s$ is the square of the c.m.\,energy and $Q_c=2/3$.
We consider the ratio~\citebk{Novikov1}
\begin{equation}
r_n 
\, \equiv \,
\frac{M_n}{M_{n-1}}
\label{mr}
\end{equation}
and extract the running gluon condensate as a function of $n$ from the
equality of the theore\-ti\-cal ratio using Eq.\,(\ref{momdef1}) and
the ratio based on Eq.\,(\ref{momdef2}) determined from experimental
data. 

For the experimental moments we use the compilation presented in
Ref.\,\citebk{Kuhn1}, where the spectral function is split into
contributions from the charmonium resonances, the charm threshold
region, and the continuum. For the latter the authors of
Ref.\,\citebk{Kuhn1} used perturbation theory since no experimental data
are available for the continuum region. We have assigned a 10\% error
for the spectral function in the continuum region. For the purely perturbative contribution of the theoretical moments we
used the compilation of analytic O$(\alpha_s^2)$ results from
Ref.\,\citebk{Kuhn1} and adopted the $\overline{\mbox{MS}}$ mass
definition $\overline{m}_c(\overline{m}_c)$ (for any renormalization
scale $\mu$). We have used the one-loop expression for the 
Wilson coefficient of the gluon condensate, and we 
have checked that for $n\le 8$ the perturbative O$(\alpha_s^2)$ corrections 
do not exceed 50\% of the O$(\alpha_s)$ corrections for $\mu$ between $1$ and $4$~GeV.  

\begin{figure}[t!] 
\begin{center}
\leavevmode
\epsfxsize=3.5cm
\epsffile[260 430 420 720]{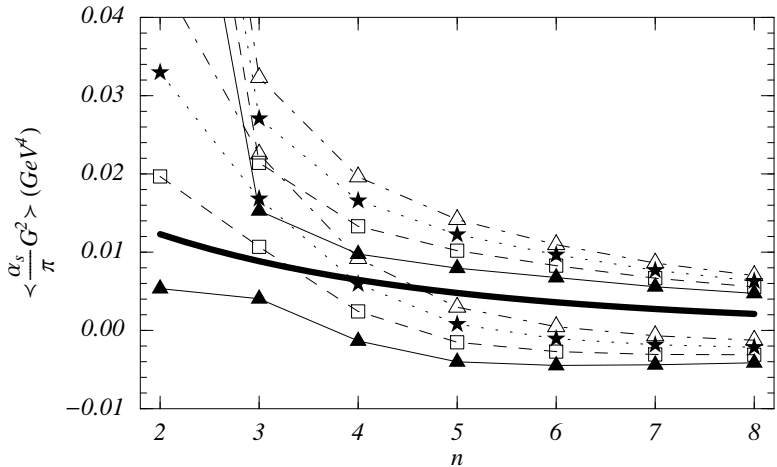}
%
%
\vspace{0cm}
 \caption{\label{figcharmonium} 
The running gluon condensate as a function of $n$ when extracted from
the ratio of moments $r_n$ for 
$\overline{m}_c(\overline{m}_c)=1.23$ (white triangles), $1.24$ 
(black stars), $1.25$ (white squares) and $1.26$~GeV (black
triangles). The area between the upper and lower symbols represent the
uncertainties. 
The thick solid line indicates the running gluon condensate as it is 
obtained from the lattice-inspired model for $g(x)$.)
}
 \end{center}
\end{figure}

Our result for the running gluon condensate as a function of
$n$ is shown in Fig.\,\ref{figcharmonium}. 
For each value of $\overline{m}_c(\overline{m}_c)$ and $n$
the area between the upper and lower symbols represents the
uncertainty due to the experimental errors of the moments and variations of 
$\alpha_s(M_Z)$ between $0.116$ and $0.120$ and of $\mu$ between $1$
and $3$~GeV. 
We see that the running gluon condensate appears to be a decreasing
function of $n$. Since the exact width of the
short-distance function $f$ for the moments is unknown we estimate it
using physical arguments. For large $n$, i.e.\,\,in the nonrelativistic
regime, the width is of the order of the quark c.m.\,kinetic energy 
$mv^2$, which scales like $m/n$ because the average quark velocity in
the $n$-th moment scales like $1/\sqrt{n}$.\,\citebk{Hoang1}
For small $n$ the relevant short-distance scale
is just the quark mass. Therefore, we take $\Omega=2m_c/n$ as the most
appropriate choice of the resolution scale. In
Fig.\,\ref{figcharmonium} the running gluon condensate is displayed for
$\lambda_A^{-1}=0.7$~GeV and, exemplarily, $A_0=0.04$ (i.e. 
$\langle(\alpha_s/\pi)G^2\rangle_{\rm lat}
(\infty)=0.024\,\,\mbox{GeV}^4$)
and for $\Omega=(2.5\,\,\mbox{GeV})/n$ as the thick solid line. We
find that the $n$-dependence of the lattice inspired running gluon
condensate and our fit result from the charmonium sum rules are
consistent. However, the uncertainties of our extraction are still quite
large, particularly for $n=2$, where the sensitivity to the error in
the continuum region is enhanced. 
We also note that the values for the running gluon condensate tend to 
decrease with increasing charm quark mass. For
$\overline{m}_c(\overline{m}_c)=1.27$~GeV we find that 
$\langle(\alpha_s/\pi)G^2\rangle$ ranges from $-0.01\,\,\mbox{GeV}^4$
to $+0.01\,\,\mbox{GeV}^4$ and shifts further towards negative values
for $\overline{m}_c(\overline{m}_c)\gsim 1.28$~GeV for all $n>2$.
Assuming the reliability of the DOE as well as the local OPE in
describing the nonperturbative effects in the charmonium sum rules
and that the Euclidean scalar functions $D(x)$ and $D_1(x)$ in the parametrization 
(\ref{parfsc}) are positive definite, our result disfavors 
$\overline{m}_c(\overline{m}_c)\gsim 1.28$~GeV.

\subsection{A model calculation}

Among the early applications of the OPE in QCD was the analysis of
non\-per\-tur\-ba\-tive effects in heavy quarkonium
systems\,\citebk{Voloshin1,Leutwyler1,Voloshin2}. Heavy 
quarkonium systems are non\-re\-la\-ti\-vis\-tic quark-antiquark bound
states for which there is the following hierarchy of the relevant  
physical scales $m$ (heavy quark mass), $m v$
(relative momentum), $m v^2$ (kinetic energy) and $\Lambda_{\rm QCD}$:
\begin{equation}
m \, \gg \, m v \, \gg \, m v^2 \, \gg \, \Lambda_{\rm QCD}
\,.
\end{equation}
Thus the spatial size of the quarkonium system $\sim (m v)^{-1}$ is
much smaller than the typical dynamical time scale $\sim (m
v^2)^{-1}$ effectively rendering the problem 1-dimensional. 
In this section we demonstrate the DOE for the
nonperturbative corrections to the $n^{\,2s+1}L_j=1^{\,3}S_1$ ground
state. We adopt the local version of the multipole expansion (OPE) for
the expansion in the ratios of the scales $m$, $m v$ and $m v^2$. The
resolution dependent expansion (DOE) is applied with respect to the
ratio of  the scales $m v^2$ and $\Lambda_{\rm QCD}$.  
The former expansion amounts to the usual treatment of the dominant
perturbative dynamics by means of a nonrelativistic two-body
Schr\"odinger equation. The interaction with the
nonperturbative vacuum is accounted for by two insertions of the local 
$\bmx\bmE$ dipole operator, $\bmE$ being the chromoelectric
field\,\citebk{Voloshin1}. The chain of VEV's of the two gluon operator
with increasing numbers of covariant derivatives 
times powers of quark-antiquark octet propagators\,\citebk{Voloshin1},
i.e. the expansion in  $\Lambda/m v^2$, is treated using the DOE.

At leading order in the local multipole expansion with respect to the
scales $m$, $m v$, and $m v^2$ the expression for the nonperturbative
corrections to the ground state energy reads
\begin{equation}
E^{np} 
\, = \,
\int_{-\infty}^\infty\! dt \, f(t) \, g(t)
\,,
\label{Enonpertdef}
\end{equation}
where
\eab
f(t) \, &=&  \, \frac{1}{36}\,
\int \frac{dq_0}{2\pi}\,\e^{i q_0(it)}\,
\int\! d^3\bmx \int\! d^3\bmy\,
\phi(x)\,(\bmx\bmy)\,
G_O\bigg(\bmx,\bmy,-\frac{k^2}{m}-q_0\bigg)\,\phi(y)
\,,\nonumber\\ 
&&
\label{Enonpertfdef}
\eae
with
\begin{eqnarray}
G_O\bigg(\bmx,\bmy,-\frac{k^2}{m}\bigg)
& = &
\sum_{l=0}^\infty\,(2l+1)\,P_l\bigg(\frac{\bmx\bmy}{x y}\bigg)\,
G_l\bigg(x,y,-\frac{k^2}{m}\bigg)
\,,
\nonumber\\[2mm]
G_l\bigg(x,y,-\frac{k^2}{m}\bigg) 
& = &
\frac{m k}{2\pi}\,(2kx)^l\,(2ky)^l\,\e^{-k(x+y)}\,\nonumber\\ 
&&\sum_{s=0}^\infty \,\frac{L_s^{2l+1}(2kx)\,L_s^{2l+1}(2ky)\,s!}
          {(s+l+1-\frac{m\,\alpha_s}{12\,k})\,(s+2l+1)!}
\,,
\nonumber\\[2mm]
\phi(x) & = & \frac{k^{3/2}}{\sqrt{\pi}}\,\e^{-k x}\,,\ \ \ \ 
k = \frac{2}{3}\,m\alpha_s\,.
\end{eqnarray}
The term $G_O$ is the quark-antiquark octet
Green-function~\citebk{Voloshin2},
and $\phi$ denotes the ground state wave function. The functions $P_n$ and
$L_n^k$ are the Legendre and Laguerre polynomials, respectively. We note
that $t$ is the Euclidean time. This is the origin of the term $it$
in the exponent appearing in the definition of the function $f$. 
\begin{figure}[t!] 
\begin{center}
\leavevmode
\epsfxsize=3.5cm
\epsffile[260 430 420 720]{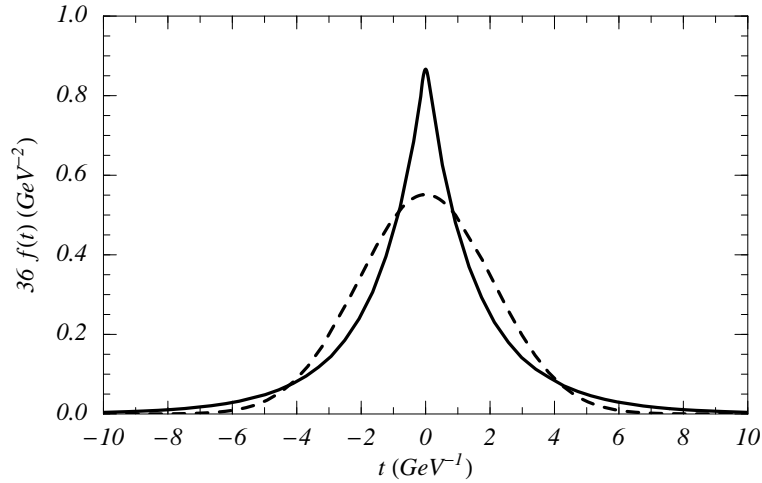}
%
%
\vspace{0cm}
 \caption{\label{figffunction} 
The perturbative short-distance function $f(t)$ for $m=5$~GeV and
$\alpha_s=0.39$ (solid line) and the leading term in the delocalized
multipole expansion of $f(t)$ for $\Omega=k^2/m$ (dashed line). 
}
 \end{center}
\end{figure}
In Fig.\,\ref{figffunction} the function $f(t)$ (solid line) is
displayed for $m=5$\,GeV and $\alpha_s=0.39$. Since the spatial
extension of the quarkonium system is neglected and the average time
between interactions with the vacuum is of the order of the inverse
kinetic energy, the characteristic width of $f$ is of order 
$(m v^2)^{-1}\sim(m\alpha_s^2)^{-1}$. As a 
comparison we have also displayed  the function 
$[\int dt^\prime f(t^\prime)\tilde e_0^\Omega(t^\prime)]
e_0^\Omega(t)$ for $\Omega=k^2/m$ (dashed line), which is the
leading term in the delocalized multipole expansion of $f$. The values
of the first few local multipole moments $f_n(\infty)$, which
correspond to the local Wilson coefficients, read
\begin{equation}
\begin{array}{ll}
f_0(\infty) \, = \, 1.6518\,\frac{m}{36\,k^4}\,, \qquad 
&  
f_2(\infty) \, = \, 1.3130\,\frac{m^3}{36\,k^8}\,, 
\\[3mm]
f_4(\infty) \, = \, 7.7570\,\frac{m^5}{36\,k^{12}}\,,  
&  
f_6(\infty) \, = \, 30.492\,\frac{m^7}{36\,k^{16}}\,, 
\\[3mm]
f_8(\infty) \, = \, 4474.1\,\frac{m^9}{36\,k^{20}}\,,
&  
f_{10}(\infty) \, = \,262709.3\,\frac{m^{11}}{36\,k^{24}}\,,\ldots\,.
\end{array}
\label{fns}
\end{equation}
The term $f_0(\infty)$
agrees with Ref.\,\citebk{Leutwyler1,Voloshin2} and $f_2(\infty)$ with
Ref.\,\citebk{Pineda}. The 
results for $f_{n>2}(\infty)$ are new. 

Let us compare the local expansion of $E^{np}$ with the
resolution-dependent expansion using the basis functions of 
Eq.\,(\ref{softb}). For the nonperturbative gluonic field strength
correlator we use a lattice inspired model of the form
\begin{eqnarray}
g(t) & = & 12\,A_0 \,\exp\Big(-\sqrt{t^2+\lambda_A^{2}}/\lambda_A + 1\Big)
\,,
\nonumber\\[2mm]
A_0 & = & 0.04~\mbox{GeV}^{4}\,,
\qquad
\lambda_A^{-1} \, = \, 0.7~\mbox{GeV}
\,,
\label{gmodel}
\end{eqnarray}
This model has an exponential large-time behavior 
and a smooth behavior for small $t$. The 
local dimension $4$ gluon condensate in this model is
\begin{equation}
\Big\langle\frac{\alpha_s}{\pi}\,G_{\mu\nu}^a G_{\mu\nu}^a\Big\rangle
\, = \, 
\frac{6\,A_0}{\pi^2} 
\, = \,  0.024\,\,\mbox{GeV}^4
\,.
\end{equation}
In Tab.\,\ref{tabquarkonium} we have displayed the exact result and
the first four terms of the resolution-dependent expansion of $E^{np}$ 
for the quark masses $m=5,25,45,90,175$~GeV and for $\Omega=\infty$
and $\Omega=k^2/m$. For each value of the quark mass the strong coupling has
been fixed by the relation $\alpha_s = \alpha_s(k)$. 
\begin{table}[t!]  
\begin{center}
\tbl{Nonperturbative corrections to the heavy quarkonium ground state level
at leading order in the multipole expansion with respect to the scales
$m$, $mv$ and $mv^2$ for various quark masses $m$ based on the model
in Eq.\,(\ref{gmodel}). Displayed are the exact result and the first
few orders in the DOE for $\Omega=\infty$ and $\Omega=k^2/m$.
The numbers are rounded off to units of $0.1$, $0.01$ or $0.001$~MeV. 
}
{\footnotesize
\begin{tabular}{|c|c|c|c||r|r|r|r|r|} \hline
\multicolumn{5}{|c|}{} & \multicolumn{2}{|c|}{$\Omega=\infty$} 
                  & \multicolumn{2}{|c|}{$\Omega=k^2/m$}
 \\ \hline
$m$ &  & $k^2/m$ & $E^{np}$ & & $f_ng_n$ & $\sum^n_{i=0}f_ig_i$ 
                              & $f_ng_n$ & $\sum^n_{i=0}f_ig_i$
 \\ 
(GeV) & \raisebox{1.5ex}[-1.5ex]{$\alpha_s$} & (MeV) & (MeV)
   & \raisebox{1.5ex}[-1.5ex]{$n$} & (MeV) & (MeV) & (MeV) & (MeV)
 \\ \hline\hline
$5$ &   $0.39$ & $0.338$ & $24.8$ & 
           0 & $38.6$      &  $38.6$    & $24.2$    & $24.2$
\\ \cline{5-9}
& & & &    2 & $-65.7$     & $-27.2$    & $-3.9$    & $20.3$
\\ \cline{5-9}
& & & &    4 & $832.7$     & $805.5$    & $12.1$    & $32.4$
\\ \cline{5-9}
& & & &    6 & $-35048.0$  & $-34242.4$  & $-43.1$  & $-10.8$
\\ \hline\hline
$25$ &  $0.23$ & $0.588$ & $12.6$ & 
           0 & $16.0$    &  $16.0$    &  $12.8$ &  $12.8$
\\ \cline{5-9}
& & & &    2 & $-9.0$    &  $7.0$     &  $-1.2$ &  $11.5$
\\ \cline{5-9}
& & & &    4 & $37.8$    &  $44.8$    &  $2.6$  &  $14.1$
\\ \cline{5-9}
& & & &    6 & $-526.6$  &  $-481.8$  &  $-6.7$ &  $7.4$
\\ \hline\hline
$45$ &  $0.19$ & $0.722$ & $4.9$ & 
           0 & $5.9$    &  $5.9$    &  $5.0$   &  $5.0$
\\ \cline{5-9}
& & & &    2 & $-2.2$   &  $3.7$    &  $-0.4$  &  $4.6$
\\ \cline{5-9}
& & & &    4 & $6.1$    &  $9.8$    &  $0.7$   &  $5.3$
\\ \cline{5-9}
& & & &    6 & $-56.4$  &  $-46.6$  &  $-1.4$  &  $3.8$
\\ \hline\hline
$90$ &  $0.17$ & $1.156$ & $1.05$ & 
           0 & $1.15$   &  $1.15$  &  $1.07$   &  $1.07$
\\ \cline{5-9}
& & & &    2 & $-0.17$  &  $0.98$  &  $-0.04$  &  $1.02$
\\ \cline{5-9}
& & & &    4 & $0.18$   &  $1.16$  &  $0.04$   &  $1.07$
\\ \cline{5-9}
& & & &    6 & $-0.65$  &  $0.51$  &  $-0.07$  &  $1.00$
\\ \hline\hline
$175$ & $0.15$ & $1.750$ & $0.245$ & 
           0 & $0.258$   &  $0.258$  &  $0.249$   &  $0.249$
\\ \cline{5-9}
& & & &    2 & $-0.016$  &  $0.242$  &  $-0.005$  &  $0.244$
\\ \cline{5-9}
& & & &    4 & $0.008$   &  $0.250$  &  $0.003$   &  $0.247$
\\ \cline{5-9}
& & & &    6 & $-0.012$  &  $0.237$  &  $-0.003$  &  $0.244$
\\ \hline
\end{tabular}}
\label{tabquarkonium}
\end{center}
\vskip 3mm
\end{table}
Note that the series are all asymptotic, i.e.\,\,there is no 
convergence for any resolution. The local expansion ($\Omega=\infty$)
is quite badly behaved 
for smaller quark masses because for $k^2/m\lsim\lambda_A^{-1}$ any
local expansion is meaningless. In particular, for $m=5$~GeV the
subleading dimension $6$ term is already larger than the parametrically
leading dimension $4$ term. This is consistent with the size of the
dimension $6$ term based on a phenomenological estimate of the local
dimension $6$ condensate. 
For quark masses, where $k^2/m\gsim\lambda_A^{-1}$, the local
expansion is reasonably good. However, for finite resolution
$\Omega=k^2/m$, the size of higher order corrections is considerably smaller
than in the local expansion for all quark masses, and the series
appears to be much better behaved. The size of the order $n$ term is
suppressed by approximately a factor $2^{-n}$ as compared to the order
$n$ term in the local expansion. We find explicitly that terms in
the series with larger $n$ decrease more quickly for finite resolution
scale as compared to the local expansion. One also observes that even
in the case $k^2/m < \lambda_A^{-1}$, where the leading term of the
local expansion overestimates the exact result, the leading
term in the delocalized expansion for $\Omega=k^2/m$ agrees with the 
exact result within a few percent. It is intuitively clear that this
feature is a general property of the delocalized expansion.

We would like to emphasize the above calculation is not intended
to provide a phenomenological determination of nonperturbative
corrections to the heavy quarkonium ground state energy level, but
rather to demonstrate the DOE within a specific model. For a realistic 
treatment of the nonperturbative contributions in the heavy quarkonium
spectrum a model-independent analysis should be carried out. In
addition, higher orders in the local multipole expansion with
respect to the  ratios of scales $m$, $m v$ and $m v^2$ should be
taken into account, which have been neglected here. These corrections
might be substantial, in particular for smaller quark masses. However,
having in mind an application to the bottomonium spectrum, we believe
that our results for the expansion in $\Lambda/mv^2$ indicate that
going  beyond the leading term in the OPE for the bottomonium ground
state is probably meaningless and that calculations based on the DOE
with a suitable choice of resolution are more reliable.

\section{Summary and Outlook}

In this talk I have presented a generalization of the expansion of 
the correlator of gauge invariant currents 
into local operator averages. It is built on 
appropriate choices of basis sets spanning 
the dual spaces which correspond to the interplay between 
perturbative and nonperturbative $N$-point correlations in 
the Euclidean propagation of a gauge invariant current. 
It was demonstrated that this framework, when applied to the $V+A$ and 
charmonium systems, yields an experimentally confirmed 
running gluon condensate. In a model calculation for the nonperturbative 
corrections to the ground state of a heavy quarkonium superior 
convergence properties of the DOE were obtained when compared 
with the OPE. 

The framework has a wealth of potential applications. For example, it can 
be used to improve the scaling relation for decay constants of heavy-light 
mesons as it is obtained in HQET (see \citebk{HoHo} for an attempt). 
The issue of the violation of local quark-hadron duality by the OPE can be 
re-addressed \citebk{Hofmann2}, and a scheme can be 
developed to estimate duality violation effects. This is of particular 
importance for the determination of standard model parameters from 
mixing and decay of $B$ mesons \citebk{Nierste}.   

\section*{Acknowledgments} 

I would like to thank the organizers for providing the atmosphere 
for a very stimulating conference. Helpful comments on the manuscript by A. Hoang 
are gratefully acknowledged.

\end{document}